\documentclass[aps,reprint,pra,showpacs,amsmath,floatfix]{revtex4-1}

\usepackage{graphicx}

\begin{document}

\title{Observing controlled state collapse in a single mechanical oscillator via a direct probe of energy variance}

\author{A. A. Gangat}
\email{a.gangat@physics.uq.edu.au}
\affiliation{ARC Centre for Engineered Quantum Systems, School of Mathematics and Physics, The University of Queensland, St Lucia, QLD 4072, Australia}

\begin{abstract}
Due to their central role in our classical intuition of the physical world and their potential for interacting with the gravitational field, mechanical degrees of freedom are of special interest in testing the non-classical predictions of quantum theory at ever larger scales.  The projection postulate of quantum theory predicts that, for certain types of measurements, continuously measuring a system induces a stochastic collapse of the state of the system toward a random eigenstate.  To date, no proposals have been made to directly observe this progressive state collapse in a mechanical oscillator.  Here we propose an optomechanical scheme to observe this fundamental effect in a vibrational mode of a mechanical membrane.  The observation in the scheme is direct (it is not inferred via an \textit{a priori} assumption of the projection postulate for the mechanical mode), and is made possible through the unprecedented feature of a direct \textit{in-situ} probe of the mechanical energy variance.  In the scheme, quantum theory predicts that a steady-state is reached as the measurement-induced collapse is counteracted by dissipation to the unmonitored environment.  Numerical simulations show this to result in a monotonic decrease in the time-averaged energy variance as the ratio of continuous measurement strength to dissipation is increased.  The measurement strength in the proposed scheme is tunable \textit{in situ}, and the behavior predicted by the simulations therefore implies a way to verifiably control the time-averaged variance of a mechanical wave function over the course of a single quantum trajectory.  The scheme's unique ability to directly probe the energy variance of the mechanical mode may also enable novel investigations of the effects on the mechanical state of coupling the mechanical mode to other quantum systems.

\end{abstract}

\maketitle

\section{Background and Overview} 

Quantum theory, whose predictions are manifest at microscopic scales, contains no intrinsic prohibition for application to macroscopic degrees of freedom.  The manifestly classical nature of the macroscopic world, however, makes such an extrapolation far from trivial in significance.  To be clear, macroscopic phenomena such as superconductivity and crystal structure were well understood to be direct manifestations of quantum mechanics long ago.  However, the possibility of a macroscopic degree of freedom (i.e. one that is collective in many microscopic degrees of freedom), such as the center of mass of a crystal, itself exhibiting a classically forbidden state or dynamical trajectory was mere speculation for several decades after the advent of quantum theory (Schr\"{o}dinger's cat is iconic of this).  This changed with the theoretical investigations of A. J. Leggett \cite{Leggett}, which provided momentum to a series of experiments in the 1980's with collective electronic degrees of freedom in superconducting circuits.  These investigations culminated in the landmark 1988 experiment of J. Clarke et al. \cite{Clarke}, which provided the first unambiguous demonstration of the quantum tunneling of a macroscopic degree of freedom (in this case, the phase difference across a Josephson junction).  Since then, microwave cavity states \cite{Brune}, C$_{60}$ molecules \cite{Zeilinger}, macroscopic currents \cite{SQUID1, SQUID2}, and even a macroscopic mechanical dilation mode \cite{OConnell}, have all been demonstrated to occupy superpositions of macroscopically distinct states, clearly validating the Schr\"{o}dinger equation for macroscopic degrees of freedom.

Quantum theoretical predictions, however, are sharply distinct from classical ones not only by way of the Schr\"{o}dinger equation, which dictates the behavior of an unmeasured system, but also through the projection postulate, which applies in the scenario of measurement.  In the case of real finite-strength quantum measurements, in which only partial information of an observable is extracted, the projection postulate predicts that a measurement will result in a partial, stochastic modification of the quantum state rather than a complete collapse \cite{Wiseman}.  Observing this fundamental effect requires a special class of measurements referred to as quantum non demolition (QND):  a QND measurement of an observable, which is possible for observables that commute with the system Hamiltonian, leaves the post-measurement state stationary (under the system Hamiltonian) in the eigenbasis of that observable, and the difference between the pre- and post-measurement states in this basis can therefore be attributed solely to the effect of measurement.  QND measurements have in fact been used to successfully observe such measurement-induced non-unitary quantum state evolution in the macroscopic degrees of freedom of microwave cavities and superconducting qubits.  In \cite{Guerlin}, successive QND measurements on an initial coherent state of a microwave cavity were used to infer the progressive collapse of the coherent state toward a nearly pure Fock state.  Complete and permanent collapse to a pure state, however, is never achievable in such a scenario due to unavoidable finite coupling to the unobserved environment; instead, if measurement is continued after the collapse process, quantum jumps between nearly pure Fock states arise, and these were also observed in the same experiment.  In \cite{Katz, Hatridge, Murch} the non-unitary modifications of a superconducting qubit state due to QND measurements were observed, and the study in \cite{Riste} observed the progressive effect of continuous measurement on the combined state of two qubits.  Quantum jumps between the ground and excited states of a superconducting qubit were first observed in \cite{Vijay}.  Regarding macroscopic mechanical degrees of freedom, however, no such experimental tests of non-unitary state evolution due to measurement have been performed.  Various theoretical investigations \cite{Santamore1, Santamore2, Martin, Jacobs1, Jacobs2, Jayich, Buks, Yanbei, Woolley, Heinrich, Gangat, Ludwig} have been done regarding proposals to observe, via continuous QND measurement, quantum jumps between nearly pure mechanical Fock states, but no proposals exist in the mechanical realm for experimental studies of the non-unitary state collapse process itself.

\begin{figure}
\includegraphics[scale=0.7]{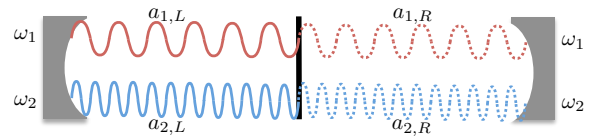}
\caption{(color online).  A depiction of the membrane-in-the-middle system wherein the mechanical membrane (black) is suspended in a cavity (grey) orthogonal to its axis.  The $a_{1,L/R}$ (red, solid/dashed) and $a_{2,L/R}$ (blue, solid/dashed) optical modes shown with respective frequencies $\omega_1$ and $\omega_2$ arise on each side of the membrane if it is perfectly opaque and still.  Due to the finite transparency, however, the normal modes $a_{1,\pm}=\frac{1}{\sqrt{2}}(a_{1,L}\pm a_{1,R})$ and $a_{2,\pm}=\frac{1}{\sqrt{2}}(a_{2,L}\pm a_{2,R})$ become the physically relevant modes.  The finite transparency of the membrane lifts the degeneracy between $a_{j,+}$ and $a_{j,-}$, and the optomechanical interaction introduces a modulation of the spectra that is proportional to $x^2$ for $a_{1,\pm}$ and $x^4$ for $a_{2,\pm}$.  $\omega_1\neq\omega_2$ such that all normal modes may be driven and monitored independently.}
\label{fig_mim}
\end{figure}

In this theoretical work we propose a scheme to observe the measurement-induced progressive collapse of a mechanical wave function in the energy eigenbasis by directly monitoring its time-averaged variance \textit{in-situ}.  The scheme is based on the platform of optomechanics (see \cite{r1,r2,r3,r4,r5,r6,r7,r8} for reviews), wherein optical field modes are coupled to the motion of mechanical resonators.  The particular system considered is the "membrane-in-the-middle" optomechanical system \cite{Thompson}, which consists of a dielectric membrane suspended in the middle of an optical cavity and orthogonal to the cavity axis (see Fig. (\ref{fig_mim})).  Depending on the equilibrium position $x_0$ of the membrane along the cavity axis, the system can exhibit a modulation of the energy of a full cavity optical mode (having annihilation operator $a$) that could be either linear or quadratic in the mechanical membrane displacement $x$ along the cavity axis from $x_0$:  $H\propto xa^{\dagger}a$ or $H\propto x^2a^{\dagger}a$.  In the quantum regime and under the rotating wave approximation, the latter becomes $H\propto b^{\dagger}ba^{\dagger}a$ for a single mechanical mode (with annihilation operator $b$), thereby providing a channel for continuous QND measurement of mechanical energy: if mode $a$ is continuously driven with a fixed drive, the phase of its continuous output signal will depend on the mechanical energy, but mode $a$ does not exchange any quanta with mode $b$ and therefore does not perturb it.  The linear in $x$ coupling provides a channel for actively cooling the mechanical mode to low occupation numbers \cite{sidebandcooling, sidebandcooling2}: driving mode $a$ at the red mechanical sideband induces a net up-conversion of the pumped photons via absorption of mechanical quanta.  Further, tilting the membrane with respect to the cavity axis can change the optomechanical modulation of select optical spectra from $x^2$ to $x^4$ at lowest order in $x$ \cite{Sankey}, providing a channel for QND measurements of $b^{\dagger}b+(b^{\dagger}b)^2$.  Also, an examination of the full optical spectra of the system reveals that the $x$, $x^2$, and $x^4$ couplings may all be achieved simultaneously with independent optical channels.  We show below that the capability of simultaneous $x$, $x^2$, and $x^4$ optomechanical coupling permits direct observation of the time-averaged variance (in the energy eigenbasis) of the quantum state of the mechanical mode while it is coupled to a thermal bath but actively cooled to the single quanta regime.

The energy variance measurement scheme requires only an \textit{a priori} assumption that the optical and mechanical modes, as well as their interaction, obey the Schr\"{o}dinger equation, and that the Born rule applies to the optical modes.  (The fully quantum nature of optical fields is well established by countless experiments, and the Schr\"{o}dinger equation was first validated for a micron-scale mechanical degree of freedom in the experiment of \cite{OConnell}.  The validity of the Schr\"{o}dinger equation for optomechanical interactions was established in the recent experiment of \cite{Palomaki}, where the interaction was used to verifiably generate entanglement between a propagating microwave field and a micromechanical oscillator.)  The projection postulate, thus far unvalidated for macroscopic mechanical degrees of freedom, predicts that the proposed scheme also permits \textit{in situ} control of the time-averaged mechanical energy variance.  As the measurement of the mechanical energy variance in the scheme does not entail an \textit{a priori} assumption of the projection postulate, it serves as a legitimate test of this prediction.  This predicted control of the energy variance is due to the interplay between a finite collapse rate of the quantum state (due to continuous QND measurement) and a finite broadening rate (due to continuous dissipation): the continuous QND measurement of $b^{\dagger}b$ through the $x^2$ optomechanical coupling produces the action of collapsing the quantum state toward a single (random) Fock state, while the coupling of the mechanical mode to dissipative channels induces a broadening of the quantum state toward a thermal state.  The steady-state between these two competing processes yields a finite time-averaged variance.  Increasing the measurement strength on $b^{\dagger}b$, which may be done \textit{in-situ} by increasing the drive strength on the optical mode coupled to $x^2$, results in a smaller steady-state time-averaged variance because the collapse rate is thereby increased.  Simultaneously, the information from the $x^4$ measurement channel may be combined with that from the $x^2$ channel to provide a direct observation of the steady state time-averaged variance.  Thus, the collapse of a mechanical quantum state in the energy eigenbasis may be observed in a single time-averaged quantum trajectory by incrementally increasing the measurement strength after each sufficiently long time-average of the measurement signals.  Though the interplay of measurement-induced collapse and dissipation-induced broadening of the mechanical quantum state in this system was conceptually understood in a previous theoretical study that involved only measurements on $b^{\dagger}b$ \cite{Gangat}, a proposal to experimentally observe and control this interplay over a range of relative strengths is unique to the present work.  The simultaneous mechanical mode cooling through the $x$ coupling serves the purpose of lowering the \textit{effective} bath temperature of the mode, thereby reducing the $x^2$ coupling strength required to substantially collapse the quantum state.

The measurement-based collapse scheme outlined above should be contrasted with the fact that, for the Hamiltonian (Eq. (\ref{model})) of the system plus its environment, the Schr\"{o}dinger equation by itself requires the mechanical quantum state to be a thermal state with a variance that slightly increases, rather than decreases, with increasing measurement strength (optical mode drive strength) \cite{varincrease}.

It is also important to note that the previous state collapse investigations with microwave cavities and superconducting qubits mentioned above were all in the regime of sufficiently efficient measurements and negligible environmentally-induced decay such that a significant fraction of the purity of the initial state was maintained over each quantum trajectory during the collapse process.  By contrast, the scheme proposed here uniquely deals with macroscopic non-unitary quantum effects in the opposite regime of significant environmental dissipation.  And, although the state of the system in such a regime is highly decohered, the decoherence arises due to entanglement with the unmonitored environment, and the state is therefore still distinctly quantum and may not be interpreted classically \cite{Schlosshauer}.  

\section{Model}To derive the model Hamiltonian for the scheme, we first follow some of the analysis of \cite{Yanbei,Ludwig} for the case of $x^2$ optical spectrum modulation in the case of a two-sided cavity.  In the membrane-in-the-middle system, when the membrane is orthogonal to the cavity axis the finite optical transmittance of the membrane and the finite optomechanical coupling give rise to the following Hamiltonian at select values of $x_0$ and valid for small $x$ of a single mechanical mode: 
\begin{align}
\tilde{H}_1= ~&\hbar \omega_{1}a_{1L}^{\dagger}a_{1L} +\hbar \omega_{1}a_{1R}^{\dagger}a_{1R} -\hbar J_1(a_{1L}^{\dagger}a_{1R} + \textrm{H.c.}) \nonumber \\& - \hbar g_{1}(x/x_{\textrm{zpf}})(a^{\dagger}_{1L}a_{1L}-a^{\dagger}_{1R}a_{1R}),
\end{align}
where $J_1$ is proportional to the transmittance of the membrane for $a_{1L/1R}$, $g_1$ is an optomechanical coupling constant, $x_{\textrm{zpf}}$ is the zero point fluctuations of the mechanical mode, and $a_{1L/1R}$ are one of the cavity mode pairs, degenerate in frequency but having different spatial mode functions, that would arise on the left/right of the membrane with frequency $\omega_{1}$ if $J_1=x=0$ (see Fig. (\ref{fig_mim})).  In terms of the full cavity modes $a_{1,\pm}=(a_{1L}\pm a_{1R})/\sqrt{2}$, the Hamiltonian is
\begin{align}
H_1&=H_{1}^{(0)} + H_{1}^{\textrm{(int)}}, \\
H_{1}^{(0)}&= \hbar \omega_{1-}a^{\dagger}_{1-}a_{1-} +\hbar \omega_{1+}a^{\dagger}_{1+}a_{1+}, \\
H_{1}^{\textrm{(int)}}&= -\hbar g_1(x/x_{\textrm{zpf}})(a^{\dagger}_{1+}a_{1-}+a^{\dagger}_{1-}a_{1+}), 
\end{align}
where $\omega_{1\pm}=\omega_1\mp J_1$.  Thus, though $a_{1,\pm}$ are degenerate without the transmittance and optomechanical interactions (i.e. when $J_1=g_1=0$), their presence modifies the optical spectrum of $a_{1,\pm}$ such that, in the physically relevant case of $J_1\gg g_{1}(x/x_{\textrm{zpf}})$, the transmittance lifts the degeneracy by an amount $2\hbar J_1$, and the optomechanical interaction induces a further perturbation of the spectrum that is quadratic in $x$.

We may analogously go beyond \cite{Ludwig,Yanbei} to model the case of $x^4$ spectrum modulation of the full cavity optical modes $a_{2,\pm}=(a_{2L}\pm a_{2R})/\sqrt{2}$:
\begin{align}
H_2&=H_{2}^{(0)} + H_{2}^{\textrm{(int)}}, \\
H_{2}^{(0)}&= \hbar  \omega_{2-}a^{\dagger}_{2-}a_{2-} +\hbar  \omega_{2+}a^{\dagger}_{2+}a_{2+}, \\
H_{2}^{\textrm{(int)}}&= -\hbar g_2(x/x_{\textrm{zpf}})^2(a^{\dagger}_{2+}a_{2-}+a^{\dagger}_{2-}a_{2+}),
\end{align}
where $\omega_{2\pm}=\omega_2\mp J_2$.  Analogous to the case with $H_1$, when $J_2\gg g_{2}(x/x_{\textrm{zpf}})^2$ the finite transmittance of the membrane lifts the degeneracy of $a_{2,\pm}$ by an amount $2\hbar J_2$, and the optomechanical interaction induces a further perturbation of the spectrum that is quartic in $x$.

We are interested here in the case of simultaneous $x$, $x^2$, and $x^4$ coupling for the fundamental mechanical mode $b$ so that the full Hamiltonian is given by
\begin{align}
H=\hbar\Omega b^{\dagger}b + H_{1} + H_{2} + H_{\textrm{1,drive}} + H_{\textrm{2,drive}} + H_{\textrm{diss}},
\end{align}
where $\Omega$ is the fundamental mechanical mode frequency, $H_{j,\textrm{drive}}=\hbar(\epsilon^*_ja_{j+}e^{i\omega_{j+}t} +\textrm{H.c.})$ encapsulates coherent drives of amplitude $\epsilon_j$ on optical modes $a_{j+}$, and $H_{\textrm{diss}}$ encapsulates all of the intrinsic and induced dissipation channels, including the mechanical sideband cooling bath that arise from the $x$ coupling \cite{sidebandcooling} and the dissipation from Raman scattering (see Appendix), for the relevant optical and mechanical modes.  It can be shown (see Appendix) that in a picture moving at the zeroth order optical and mechanical dynamics and after the rotating wave approximation, the model Hamiltonian becomes
\begin{align}
H_{\textrm{model}}= &-\frac{\hbar}{2}g_1^2 A_1 n_{1+} n_b -\frac{\hbar}{2}g_2^2 A_2  n_{2+} (n_b^2 + A n_b) \nonumber \\ &+ H_{1,\textrm{drive}}' + H_{2,\textrm{drive}}' + H_{\textrm{diss}},
\label{model}
\end{align} 
where $n_b=b^{\dagger}b$, $n_{j\pm}=a_{j\pm}^{\dagger}a_{j\pm}$, $H_{j,\textrm{drive}}'=\hbar(\epsilon^*_j a_{j+} +\textrm{H.c.})$, $A_1=2(\frac{1}{2J_1-\Omega}+\frac{1}{2J_1+\Omega})$, $A=B/A_2$, $A_2=\frac{4}{J_2}+\frac{2J_2}{J_2^2-\Omega^2}$, $B=\frac{4}{J_2}+\frac{2J_2+4\Omega}{J_2^2-\Omega^2}$.

\section{Collapse observation and control}The protocol for observation and control of measurement-induced mechanical quantum state collapse is as follows.  As mentioned in Section I., the projection postulate dictates that the steady-state mechanical quantum state under continuous measurement of $n_b$ is the result of a competition between collapse due to acquisition of information in the measurement record and broadening due to loss of information through the unmonitored dissipation channels.  Although in this situation the quantum state itself fluctuates in time due to the continuous QND measurement, the long time-average of its variance is constant.  If the dissipation rates are constant, increasing the measurement strength on $n_b$ results in a smaller time-averaged variance for $n_b$.  As the $n_b$ measurement strength is proportional to the drive on $a_{1+}$, the drive strength serves as an {\em in-situ} experimental knob for the time-averaged variance.  Selecting any values for the $a_{j+}$ drive strengths, one may take a long time-average of both the output of $a_{1+}$ and its squared value to respectively extract $\overline{\langle n_b \rangle_s(t)}$ and $\overline{\langle n_b\rangle_s(t)^2}$, where the subscript "$s$" denotes that $\langle u \rangle_s(t)$ is not an ensemble average but the mean value of the observable \textit{u} according to the \textit{single} mechanical quantum state at time $t$.  Simultaneously, one may obtain $\overline{\langle n_b^2\rangle_s(t)}+A\overline{\langle n_b\rangle_s(t)}$ from a long time-average of the output of $a_{2+}$ and combine it with the information from $a_{1+}$ to determine $\overline{\langle n_b^2 \rangle_s(t)}$.  Thus, one obtains sufficient information to determine the time-averaged steady-state mechanical energy variance $\overline{\sigma_b(t)^2}=\overline{\langle n_b^2 \rangle_s(t) - \langle n_b\rangle_s(t)^2}$ of the single quantum trajectory at the selected values of the drive strengths.  This experimental procedure requires no \textit{a priori} assumption of the projection postulate.  Repeating this procedure for incrementally stronger values of the drive on $a_{1+}$, one may therefore test for the collapse of $\overline{\sigma_b(t)^2}$ with increasing measurement strength as predicted by the projection postulate.  By relying on the measurements through the $a_{1+}$ channel to collapse the quantum state, this protocol accommodates the fact that the experimentally observed $g_2$ is very weak \cite{Sankey}; the information required from $a_{2+}$ can always be obtained through sufficiently long time-averages.

Being able to collapse the quantum state in this manner, however, implies certain parameter constraints.  The study in \cite{Gangat} established two fundamental conditions for ensuring that the mechanical quantum state remained collapsed to a nearly pure Fock state so that quantum jumps would arise: both $\kappa_{1+}$ (the damping rate for mode $a_{1+}$) and the $n_b$ measurement rate $\Gamma_{1}$ must be much greater than the mechanical Fock state decay rate.  The collapse observation and control protocol in the present proposal therefore requires that both $\kappa_{1+}$ and the maximum attainable value of $\Gamma_{1}$ satisfy the same constraint.  The study in \cite{Gangat} considered the special case of a one-sided cavity where coupling to a thermal bath was the only source of mechanical dissipation.  In the more realistic case that we consider here of a two-sided cavity with continuous sideband cooling of the mechanical mode, the fundamental conditions may be expressed as
\begin{align}
\Gamma_{1}^{(\textrm{max})}~\textrm{and}~ \kappa_{1+} \gg& ~\gamma_{\textrm{th}}\big[(\overline{n}_{\textrm{th}}+1)\overline{n}_b +\overline{n}_{\textrm{th}}(\overline{n}_b+1)\big] \nonumber \\ &+ (\gamma_{1,Ram} + \gamma_{2,Ram} + \gamma_{opt})\overline{n}_b,
\label{constraints}
\end{align}
where $\Gamma_{1}^{(\textrm{max})}$ denotes the maximum attainable value of $\Gamma_{1}$, $\gamma_{\textrm{th}}$ is the mechanical dissipation rate to the mechanical mode thermal bath of average occupation $\overline{n}_{\textrm{th}}$, $\gamma_{opt}$ is the mechanical dissipation rate to the zero temperature optical cooling bath induced by the sideband cooling \cite{sidebandcooling}, and $\gamma_{j,Ram}$ are due to Raman scattering processes (see Appendix).  Although $\gamma_{1,Ram}$ is itself proportional to $n_{1+}$, which must be increased to increase $\Gamma_{1}$ for each successive collapse increment of the protocol, the total mechanical cooling rate $\gamma_{\textrm{cool}}=\gamma_{1,Ram}+\gamma_{2,Ram}+\gamma_{opt}$ may be kept constant by simultaneously reducing $\gamma_{opt}$ via {\em in-situ} adjustment of the sideband cooling drive \cite{sidebandcooling}.

The contraints on $\Gamma_1^{(\textrm{max})}$ and $\kappa_{1+}$ show that the feasibility of the scheme entails that $\overline{n}_b$ be small.  From detailed balance, $\overline{n}_b=(\gamma_{\textrm{cool}}\overline{n}_{opt}+\gamma_{\textrm{th}}\overline{n}_{\textrm{th}})/(\gamma_{\textrm{cool}}+\gamma_{\textrm{th}})$.  As the optical bath occupation $\overline{n}_{opt}$ is very small, choosing $\gamma_{\textrm{cool}}\approx\gamma_{\textrm{th}}\overline{n}_{\textrm{th}}$ yields $\overline{n}_b\approx1$ and $\Gamma_{1}^{(\textrm{max})}, \kappa_{1+} \gg 4\gamma_{\textrm{th}}\overline{n}_{\textrm{th}}$.  Thus, steady-state $\overline{n}_b\approx1$ can be achieved via continuous sideband cooling that is simultaneous with the collapse measurement and quantum jump measurement protocols without significantly increasing $\Gamma_{1}^{(\textrm{max})}$ beyond what would be required in the absence of continuous sideband cooling where the mechanical mode was instead passively cooled to $\overline{n}_b\approx1$.  Observing phonon number quantum jumps and quantum state collapse with simultaneous sideband cooling may prove to be an experimentally more viable route than with passive cooling.

The authors of \cite{Yanbei} derive the additional condition $g_1^2>\kappa_{1+}\kappa_{1-}$ for detection of quantum jumps in energy, where $\kappa_{1-}$ is the damping rate for mode $a_{1-}$, by requiring that the phonon number measurement rate be greater than the mechanical Fock state decay rate due to the Raman process mentioned above.  However, because the measurement plays the dual role of detecting the Fock state and also collapsing the quantum state to create the Fock state, what is actually required is that the measurement rate be {\em much} greater than the Fock state decay rate.  This was established in \cite{Gangat} and is reflected in the constraint on $\Gamma_1^{(\textrm{max})}$ above.  The true requirement is therefore
\begin{align}
g_1^2\gg\kappa_{1+}\kappa_{1-}
\end{align}

\section{Simulations and implications for experimental signatures}In this section we produce numerical predictions that assume the projection postulate to hold true for the mechanical mode, and we present expressions for the experimental photocurrents that are derived without recourse to the projection postulate for the mechanical mode.  The experimental signals may therefore be used to test the numerical predictions.  To produce the numerical predictions, we consider the case where the measurement modes $a_{j+}$ are strongly driven so that $a_{j+}\rightarrow \alpha_j + a_{j+}'$, where $\alpha_j$ are the steady-state background amplitudes of $a_{j+}$ and $a_{j+}'$ are the quantum fluctuations on top of $\alpha_j$.  We may therefore proceed in analogy to \cite{Gangat} to move to a displaced picture for the modes $a_{j+}$ and use the following stochastic master equation (SME) \cite{Wiseman} for the (conditional) system density matrix $\rho_c$ to treat the transmitted outputs of $a_{j+}$ as continuously observed via homodyne detection and the rest of the dissipative channels as unobserved:
\begin{align}
d\rho_c=&-\frac{i}{\hbar}[H_{\textrm{I}},\rho_c]dt + \big(\gamma_{\textrm{th}}(\overline{n}_{\textrm{th}}+1)+\gamma_{\textrm{cool}}\big)\mathcal{D}[b]\rho_c dt \nonumber \\ &+ \gamma_{\textrm{th}}\overline{n}_{\textrm{th}}\mathcal{D}[b^\dag]\rho_c dt  + \sum_{j=1}^2(\kappa_{j+}+\kappa_{j+,Ram})\mathcal{D}[a_{j+}]\rho_c dt \nonumber \\ &+ \sum_{j=1}^2\sqrt{\eta\kappa_{j+,t}}dW_j\mathcal{H}[a_{j+}e^{-i\frac{\pi}{2}}]\rho_c.
\end{align}
The subscript "$c$" denotes that the quantity is \textit{conditioned} upon the measurement record, as required by the projection postulate.  Here we have dropped the prime from $a_{j+}'$ for simplicity, $H_{\textrm{I}}=-\hbar\frac{\chi_1}{2}(a_{1+}^{\dagger} + a_{1+})n_b - \hbar\frac{\chi_2}{2}(a_{2+}^{\dagger} + a_{2+})(n_b^2+An_b)$ is the linearized interaction Hamiltonian in the displaced picture, $\chi_j=2g_j^2A_j\alpha_j$, ${\cal D}[c]\rho=c\rho c^\dagger -c^\dagger c\rho/2-\rho c^\dagger c/2$ is the dissipation superoperator, $\mathcal{H}[c]\rho=c\rho + \rho c^\dag - \textrm{Tr}(c\rho + \rho c^\dag)$ is the measurement superoperator, $\eta$ is the efficiency of the detectors, and $dW_j$ are independent Wiener increments.  Each optical mode $a_{j+}$ has three dissipation channels at zero temperature with corresponding dissipation rates: reflected signal ($\kappa_{j+,r}$), transmitted signal ($\kappa_{j+,t}=\kappa_{j+}-\kappa_{j+,r}$), and Raman scattering decay ($\kappa_{j+,Ram}$) as mentioned above.  The mechanical mode $b$ has four dissipation channels: thermal bath dissipation ($\gamma_{\textrm{th}}$) and cooling dissipation ($\gamma_{\textrm{cool}}=\gamma_{1,Ram}+\gamma_{2,Ram}+\gamma_{opt}$), which consists of two Raman scattering channels ($\gamma_{1,Ram}$ and $\gamma_{2,Ram}$) and sideband cooling ($\gamma_{opt}$).   As explained in the previous section, $\gamma_{\textrm{cool}}$ may be considered constant over the entire collapse measurement process.  Under only the assumptions of the validity of the Schr\"{o}dinger equation and the Born rule for the optical modes, the homodyne measurement photocurrents may be derived as \cite{Wiseman}
\begin{align}
i_j(t)=\eta\kappa_{j+,t}\langle a_{j+}e^{-i\frac{\pi}{2}}+a_{j+}^{\dag}e^{i\frac{\pi}{2}}\rangle(t) + \sqrt{\eta\kappa_{j+,t}}\xi_j(t),
\end{align}
where the noise term $\xi_j(t)$ is due to the local oscillator and numerically is $\xi_j(t)=dW_j/dt$.

\begin{figure}
\includegraphics[scale=.45]{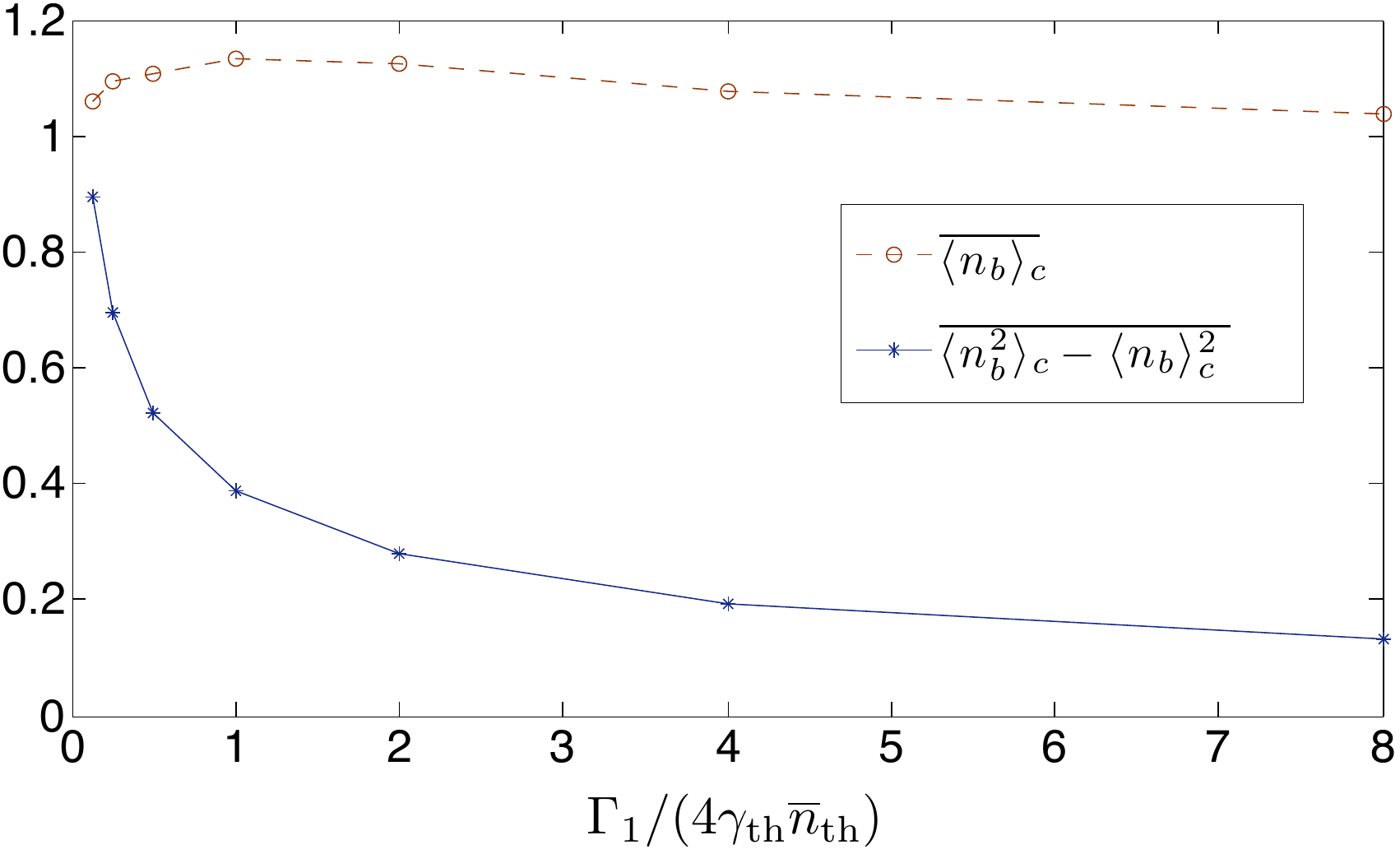}
\caption{(color online).  Simulated collapse of the quantum state in the energy eigenbasis with increasing $n_b$ measurement strength ($\Gamma_1$) and constant $n_b^2+An_b$ measurement strength ($\Gamma_2$).  For each value of $\Gamma_1$, the system is first allowed to reach steady-state, then $\overline{\langle n_b \rangle_c}$, $\overline{\langle n_b \rangle_c^2}$, and $\overline{\langle n_b^2 \rangle_c}$ are obtained by computing $\langle n_b \rangle_c$ and $\langle n_b^2 \rangle_c$ (the subscript "$c$" denotes that the quantity is \textit{conditioned} upon the measurement record, as required by the projection postulate) at each subsequent time step and averaging for a sufficiently long time.  The steady-state conditional variance  decreases monotonically as a function of $\Gamma_1$ and becomes much less than 1 when $\Gamma_1\gg4\gamma_{\textrm{th}}\overline{n}_{\textrm{th}}$, the regime in which quantum jumps are predicted to arise in $\langle n_b \rangle_c(t)$ \cite{Gangat}.  
Simulation parameters: $\Gamma_2=10^{-7}(4\gamma_{\textrm{th}}\overline{n}_{\textrm{th}})$, $\overline{n}_{\textrm{th}}=5\times10^3$, $\gamma_{\textrm{cool}}=\gamma_{\textrm{th}}\overline{n}_{\textrm{th}}$, and $A=1$.}
\label{fig_sim}
\end{figure}

As per Eq. (\ref{constraints}), $a_{1+}$ must adiabatically follow the mechanical mode energy state.  For computational simplicity we assume that $a_{2+}$ does as well so that from the quantum Langevin equations we find $a_{1+}=i\frac{\chi_1}{\kappa_{1+}+\kappa_{1+,Ram}}n_b$ and $a_{2+}=i\frac{\chi_2}{\kappa_{2+}+\kappa_{2+,Ram}}\big(n_b^2+An_b \big)$.  Using this and the fact that the mechanical mode density matrix remains diagonal due to environmental decoherence \cite{decoherence}, we find
\begin{align} 
dp_n=~&\gamma_{\textrm{th}}\overline{n}_{\textrm{th}} \big[np_{n-1}-(n+1)p_n \big]dt \nonumber \\&+ (\gamma_{\textrm{th}}(\overline{n}_{\textrm{th}}+1)+\gamma_{\textrm{cool}})\big[(n+1)p_{n+1}-np_n\big]dt \nonumber \\ 
&- 2\sqrt{\eta\Gamma_1}\big(n-\langle n \rangle_c \big)p_n dW_1\nonumber \\&- 2\sqrt{\eta\Gamma_2}\big((n^2+An)-\langle n^2 + An \rangle_c \big)p_n dW_2,
\label{SRE}
\end{align}
where $\Gamma_j=\frac{\chi_j^2\kappa_{j+,t}}{(\kappa_{j+}+\kappa_{j+,Ram})^2}$, and $p_n$ is the occupation probability of the $n^{th}$ mechanical Fock state.  The photocurrents, now under the additional assumption that the Schr\"{o}dinger equation applies to the mechanical mode and the optomechanical interaction, become
\begin{align}
i_1(t)&=2\eta\chi_1\langle n_b \rangle(t) + \sqrt{\eta\kappa_{1+,t}}\xi_1(t), \\
i_2(t)&=2\eta\chi_2\langle n_b^2+An_b \rangle(t) + \sqrt{\eta\kappa_{2+,t}}\xi_2(t).
\end{align}
Here, $\Gamma_1$ is the same phonon number measurement rate discussed in the previous section.  Experimentally, provided that $A$ (defined above) is known, for fixed values of $\Gamma_j$ sufficiently long time-averages of $i_1(t)$, $i_1^2(t)$, and $i_2(t)$ respectively yield the values of $\overline{\langle n_b \rangle}$, $\overline{\langle n_b \rangle^2}$ \cite{note2}, and $\overline{\langle n_b^2 \rangle}$.  As the derivation of the photocurrent expressions does not require an assumption of the projection postulate for the mechanical mode, the experimental acquisition of these values through the photocurrents can serve as a test of projection postulate in the mechanical realm.  The prediction of the projection postulate is that $\overline{\sigma_b}$ obtained from these experimental values will follow the monotonic behavior in Fig. \ref{fig_sim}, which is from simulations wherein the system density matrix is conditioned upon the measurement record.  We remind the reader that $\Gamma_j$ are proportional to $\alpha_j$, which can be adjusted {\em in situ} by varying the optical drive strengths.

Assuming $T\approx300$ mK, $\Omega\approx2\pi\times1$ MHz \cite{Sankey, Flowers-Jacobs}, and $\gamma_{\textrm{th}}=2\pi\times0.1$ Hz \cite{Sankey}, we find $\overline{n}_{\textrm{th}}=5\times10^3$.  As per above, we set $\gamma_{\textrm{cool}}=\gamma_{\textrm{th}}\overline{n}_{\textrm{th}}$ so that $\overline{n}_b\approx1$.  Arbitrarily setting $A=1$, we assume $\kappa_{1+}\gg4\gamma_{\textrm{th}}\overline{n}_{\textrm{th}}$ so that Eq. (\ref{SRE}) is valid and we numerically integrate it for different values of $\Gamma_1$ to produce the data shown in Fig. \ref{fig_sim}.

Finally, we note that the environment is modeled here as a bath of non-interacting harmonic oscillators, but it may be that two-level systems also play an appreciable role in the mechanical dissipation \cite{bath1,bath2}.  This should not however affect the qualitative feature of a monotonic collapse, which simply depends on the generic effect of dissipation to an unmonitored environment.

\section{Conclusion and Outlook} This work presents a feasible scheme to observe the measurement-induced collapse of a mechanical quantum state through a single time-averaged quantum trajectory.  The proposed observation does not entail an \textit{a priori} assumption of the projection postulate for the mechanical quantum state and can therefore serve as a fundamental test of it in the mechanical realm.  This is of importance for testing quantum theory at macroscopic scales.  The state of the mechanical oscillator in the absence of measurement is a thermal state, a result of entanglement with its unmonitored environment \cite{Popescu,Goldstein}, and the observable \textit{in situ} control of its variance via measurement may lead to novel applications or other fundamental tests, as it is effectively a control of the amount of entanglement shared between the mechanical mode and its unmonitored environment.  Also, setting the strengths of both measurement channels to be extremely weak can serve as a means of probing the time-averaged mechanical energy variance with very little measurement disturbance, and could serve as an \textit{in situ} means of probing the time-averaged effects on the mechanical energy variance of other quantum systems that may be coupled to the mechanical mode.

The scheme that we present is not too far from experimental reach as the latest iteration of the particular system considered shows orders of magnitude improvement in the $x^2$ optomechanical coupling strength \cite{Flowers-Jacobs}.  A further order of magnitude increase in $g_1$ and optimization of $\kappa_{1\pm}$ should achieve the requirement $g_1^2\gg\kappa_{1+}\kappa_{1-}$.

\section*{Acknowledgments} A.A.G. is grateful to Gerard J. Milburn for discussions and a critical reading of the early manuscript, and acknowledges the support of the Australian Research Council grants FF0776191 and CE110001013 and the University of Queensland.

\section*{Appendix}
Below is a derivation of the model Hamiltonian $H_{\textrm{model}}$ in the main text.  An explanation of the Raman scattering processes is contained in the final paragraph.

The full Hamiltonian is given by 
\begin{align}
H=\hbar\Omega b^{\dagger}b + H_{1} + H_{2} + H_{\textrm{1,drive}} + H_{\textrm{2,drive}} + H_{\textrm{diss}},
\end{align}
where
\begin{align}
H_1&=H_{1}^{(0)} + H_{1}^{\textrm{(int)}},\\
H_{1}^{(0)}&= \hbar \omega_{1-}a^{\dagger}_{1-}a_{1-} +\hbar \omega_{1+}a^{\dagger}_{1+}a_{1+},\\
H_{1}^{\textrm{(int)}}&= -\hbar g_1(x/x_{\textrm{zpf}})(a^{\dagger}_{1+}a_{1-}+a^{\dagger}_{1-}a_{1+}),
\end{align}
and
\begin{align}
H_2&=H_{2}^{(0)} + H_{2}^{\textrm{(int)}},\\
H_{2}^{(0)}&= \hbar  \omega_{2-}a^{\dagger}_{2-}a_{2-} +\hbar  \omega_{2+}a^{\dagger}_{2+}a_{2+}, \\
H_{2}^{\textrm{(int)}}&= -\hbar g_2(x/x_{\textrm{zpf}})^2(a^{\dagger}_{2+}a_{2-}+a^{\dagger}_{2-}a_{2+}),
\end{align}
and $H_{\textrm{1,drive}} + H_{\textrm{2,drive}} + H_{\textrm{diss}}$ is given in the main text.  If $H_j^{\textrm{(int)}}$ are treated as perturbations to $H_j^{(0)}$, the perturbations of the frequencies $\omega_{j\pm}$ can appear as a power series in $g_j$, and this is what we seek by using the approach of \cite{Ludwig}.

We first find the zeroth order in $g_j$ time-dependence of $x$ and $(a^{\dagger}_{j+}a_{j-}+a^{\dagger}_{j-}a_{j+})$ from the bare system Hamiltonian in the $a_{j,L/R}$ basis: 
\begin{align}
H_{sys}= \hbar\Omega b^{\dagger}b &+ \sum_{j} \Big( \hbar \omega_{j}a_{jL}^{\dagger}a_{jL} +\hbar \omega_{j}a_{jR}^{\dagger}a_{jR} \\
&-\hbar J_j(a_{jL}^{\dagger}a_{jR} + \textrm{H.c.}) \nonumber \\ &- \hbar g_{j}(x/x_{\textrm{zpf}})^j(a^{\dagger}_{jL}a_{jL}-a^{\dagger}_{jR}a_{jR}) \Big),
\end{align}
then plug into $H_j^{\textrm{(int)}}$ to find the force exerted on the mechanical mode:

\begin{align}
F_j(t)&=-\frac{\partial H_j^{\textrm{(int)}}}{\partial x} \nonumber \\ 
&=\frac{\hbar g_j}{x_{\textrm{zpf}}}j(b^{\dagger}e^{i\Omega t}+\textrm{H.c.})^{j-1}(a^{\dagger}_{j+}a_{j-}e^{-i2J_jt}+\textrm{H.c.})
\end{align}
For $j=1$ this induces the forced mechanical motion 
\begin{align}
x_{\textrm{f},1}&=F_1(t)/m(\Omega^2-4J_1^2), \\
p_{\textrm{f},1}&=\dot{F}_1(t)/(\Omega^2-4J_1^2),
\end{align}
while for $j=2$ the forced mechanical motion is
\begin{align}
x_{\textrm{f},2}&=\frac{1}{2}\frac{F_{2-}(t)}{m(\Omega^2-(2J_2-\Omega)^2)}+\frac{1}{2}\frac{F_{2+}(t)}{m(\Omega^2-(2J_2+\Omega)^2)}, \\
p_{\textrm{f},2}&=m\dot{x}_{\textrm{f},2},
\end{align}
where $F_{2,\pm}(t)=\frac{2\hbar g_2}{x_{\textrm{zpf}}}(a_+^{\dagger}a_-b^{\dagger}e^{-i(2J\pm\Omega)t}+\textrm{H.c.})$.  $x_{\textrm{f},j}$ and $p_{\textrm{f},j}$ are the first order in $g_j$ perturbations to the mechanical dynamics.

The next step is to make a time-dependent canonical transformation of the bare system Hamiltonian to a frame moving at the first order in $g_j$ dynamics so that it cancels and the higher order in $g_j$ perturbations become explicit:
\begin{align}
e^{i(S_1+S_2)}&\big(\hbar\Omega b^{\dagger}b + H_{1} + H_{2} \big)e^{-i(S_1+S_2)}=\hbar\Omega b^{\dagger}b + H_{1}^{(0)} \nonumber \\ &+H_{2}^{(0)} +\frac{i}{2}[S_1,H_1^{\textrm{(int)}}]+\frac{i}{2}[S_2,H_2^{\textrm{(int)}}] \nonumber \\ &+\frac{i}{2}[S_1,H_2^{\textrm{(int)}}]+\frac{i}{2}[S_2,H_1^{\textrm{(int)}}]+O(g_j^3),
\end{align}
where $S_j=x_{\textrm{f},j}p/\hbar - p_{\textrm{f},j}x/\hbar$, $x=x_{\textrm{zpf}}(b^{\dagger}+b)$, $p=i\hbar(b^{\dagger}-b)/x_{\textrm{zpf}}$, $x_{\textrm{zpf}}=\sqrt{\hbar/2m\Omega}$, and $m$ is the mechanical mode effective mass.   Thus finding the expansion of the system Hamiltonian in powers of $g_j$, we plug into the full Hamiltonian, transform to a picture moving at the zeroth order optical and mechanical dynamics, make the rotating wave approximation, and drop the first order in $g_j$ terms of the expansion (as they do not modify the optical spectra) to find the effective Hamiltonian $H_{\textrm{eff}}=H_{1,\textrm{eff}}^{(2)} + H_{2,\textrm{eff}}^{(2)} +H_{1,\textrm{drive}}' + H_{2,\textrm{drive}}' + H_{\textrm{diss}}+O(g_j^3)$, where
\begin{align}
H_{1,\textrm{eff}}^{(2)}= &\hbar \frac{g_1^2}{2}\Big( \frac{4J_1}{4J_1^2-\Omega^2} \Big) \Big( n_{1-} - n_{1+} \Big) (1+2n_b) \nonumber \\
&+ \hbar \frac{g_1^2}{2}\Big(  \frac{2\Omega}{4J_1^2-\Omega^2} \Big) \Big(2n_{1-}n_{1+} + n_{1+} + n_{1-} \Big),
\label{H1eff}
\end{align}
\begin{align}
H_{2,\textrm{eff}}^{(2)}=&\hbar\frac{g_2^2}{2} \Big(n_{2-}-n_{2+} \Big) \Big( A_2 n_b^2 + B n_b + C \Big) \nonumber \\&- \hbar g_2^2\frac{4\Omega}{J_2^2-\Omega^2} n_{2+}n_{2-}\Big(3 + 4 n_b \Big) \\& +\hbar g_2^2 \frac{\Omega^2/J_2}{J_2^2-\Omega^2}n_{2+},
\label{H2eff}
\end{align} 
$n_b=b^{\dagger}b$, $n_{j\pm}=a_{j\pm}^{\dagger}a_{j\pm}$, $H_{j,\textrm{drive}}'=\hbar(\epsilon^*_j a_{j+} +\textrm{H.c.})$, $H_{\textrm{diss}}$ is unchanged as it is modeled with Lindbladian superoperators (see below) that are invariant under transformation to the zeroth order dynamics, $A_2=\frac{4}{J_2}+\frac{2J_2}{J_2^2-\Omega^2}$, $B=\frac{4}{J_2}+\frac{2J_2+4\Omega}{J_2^2-\Omega^2}$, and $C=\frac{\Omega(3+\Omega/J_2)}{J_2^2-\Omega^2}+\frac{1}{J_2}+\frac{2J_2}{J_2^2-\Omega^2}$.  These expressions are valid for all values of the ratios $\Omega/J_j$.  We note that the coupling enhancement when $2J_1-\Omega\ll J_1,\Omega$ noted in \cite{Ludwig} for $x^2$ coupling has an analogous counterpart in the case of $x^4$ coupling when $J_2-\Omega\ll J_2,\Omega$, and if $J_j\gg\Omega$ the second lines of Eqs. (16) and (17) become negligible.

It is clear that the frequency of mode $a_{1,\pm}$ is sensitive to $n_b$ and the frequency of $a_{2,\pm}$ is sensitive to both $n_b$ and $n_b^2$.  When modes $a_{j+}$  are continuously driven, their outputs therefore yield continuous information on $n_b$ and $n_b^2$.  However, in the case that $2J_1-\Omega\ll J_1,\Omega$, the frequency of $a_{1+}$ is actually sensitive to $(n_{1-}-n_b)$.  Although we model the situation where $a_{j-}$ are left undriven, $H_j^{\textrm{(int)}}$ support secondary Raman processes whereby quanta from $a_{j+}$ combine with phonons to scatter into $a_{j-}$.  Letting $\kappa_{j\pm}$ be the total intrinsic dissipation rates of $a_{j\pm}$, reference \cite{Ludwig} determines this to occur at a rate $\gamma_{1,Ram}\overline{n}_b=\kappa_{1+,Ram}\overline{n}_{1+}=g_1^2 \overline{n}_{1+} \overline{n}_b \kappa_{1-}/(2J_1-\Omega)^2$ for scattering from $a_{1+}$ to $a_{1-}$, and an analogous golden rule calculation yields $\gamma_{2,Ram}\overline{n}_b=\kappa_{2+,Ram}\overline{n}_{2+}=g_2^2 \overline{n}_{2+} \overline{n}_b \kappa_{2-}/(2J_2-2\Omega)^2$ for scattering from $a_{2+}$ to $a_{2-}$. Therefore $\gamma_{1,Ram}$ is also amplified when $2J_1-\Omega\ll J_1,\Omega$, and a non-negligible $n_{1-}$ may result.  In the case of quantum jump measurements, although not noted in the study of \cite{Ludwig}, this Raman process can thereby register as a double jump in the output signal of $a_{1+}$ due to its sensitivity to $(n_{1-}-n_b)$.  Similarly, in the case of $J_2-\Omega\ll J_2,\Omega$, $a_{2+}$ becomes sensitive to $n_{2-}$ and $\gamma_{2,Ram}$ is amplified so that $n_{2-}$ may become non-negligible.  We are interested here, however, in the case of long time-averages of the outputs of $a_{j+}$ and assume that even in the cases where the second lines of Eqs. (\ref{H1eff}) and (\ref{H2eff}) become significant, their time-averaged contributions to the output signals may be subtracted away by, for example, independently monitoring the outputs of $a_{j-}$ \cite{aj-}.  For simplicity, then, we deal with the following model Hamiltonian:

\begin{align}
H_{\textrm{model}}= &-\frac{\hbar}{2}g_1^2 A_1 n_{1+} n_b -\frac{\hbar}{2}g_2^2 A_2  n_{2+} (n_b^2 + A n_b) \nonumber \\ &+ H_{1,\textrm{drive}}' + H_{2,\textrm{drive}}' + H_{\textrm{diss}},
\end{align} 
where $A_1=2(\frac{1}{2J_1-\Omega}+\frac{1}{2J_1+\Omega})$, $G_2=g_2^2A_2$, $A=B/A_2$, and dissipation due to the environment and Raman processes is contained in $H_{\textrm{diss}}$.


\begin{references}
\bibitem{varincrease} The $b^{\dagger}b$ measurement strength is proportional to the number of optical quanta ($n_{1+}$) in the coupled optical mode.  Increasing $n_{1+}$ decreases the effective mechanical oscillator frequency and therefore, if there is no measurement-induced collapse, should cause the mechanical energy variance to slightly increase given a thermal bath at a constant temperature.
\bibitem{decoherence} The decoherence theory of open quantum systems establishes that the density matrix of a system (here the mechanical mode) coupled to a macroscopic measurement apparatus (here the cavity mode) will be diagonal in the eigenbasis of the system observable (here the phonon number) that is coupled to the apparatus due to the environmentally-induced rapid decay of the off-diagonal elements in that basis.  In the present case, the mechanical mode density matrix will therefore be diagonal in the energy eigenbasis.  For reviews of decoherence theory, see \cite{Zurek, SchlosshauerRMP}.
\bibitem{note2} Because $\xi_1(t)$ is a Gaussian random variable with zero mean, the long time-average of $i_1(t)^2$ yields a term proportional to $\overline{\langle n_b\rangle^2}$ plus a constant that needs to be calibrated away.
\bibitem{aj-} To see this in more detail, note that when the second lines of Eqs. (\ref{H1eff}) and (\ref{H2eff}) are non-negligible, the photocurrents $i_1(t)$, $i_1^2(t)$, and $i_2(t)$ defined in the main text will include undesirable terms proportional to $\langle n_{j-} \rangle$ and $\langle n_{1-} \rangle^2$.  By continuously monitoring with photon counters the outputs of $a_{j-}$, which are spectrally distinct from one another and from $a_{j+}$, $\overline{\langle n_{j-} \rangle}$ and $\overline{\langle n_{1-} \rangle^2}$ may be directly measured and their undesirable contributions to the time-averages of $i_1(t)$, $i_1^2(t)$, and $i_2(t)$ may therefore be subtracted.
\end{references}
\end{document}